\documentstyle[epsfig]{mn}

\def\Msun{\ifmmode{~{\rm M}_\odot}\else${\rm M}_\odot$~\fi}
\def\kms{\ifmmode{$~km\thinspace s$^{-1}}\else km\thinspace s$^{-1}$\fi}

\title[K dwarfs and the chemical evolution of the Solar cylinder]{K dwarfs and
the chemical evolution of the Solar cylinder}

\author[E. Kotoneva, C. Flynn, C. Chiappini, F. Matteucci]{Eira Kotoneva$^1$,
Chris Flynn$^{2,3,}$\thanks{Research Fellow of the Academy of Finland}, Cristina Chiappini$^4$, Francesca Matteucci$^{5,4}$\\
$^1$Tuorla Observatory, Piikki\"o, FIN-21500, Finland; eianko@astro.utu.fi\\
$^2$Tuorla Observatory, Piikki\"o, FIN-21500, Finland; cflynn@astro.utu.fi\\
$^3$Centre for Astrophysics and Supercomputing, Swinburne University of
Technology, Hawthorn, Australia\\
$^4$INAF - Osservatorio Astronomico di Trieste, Via G.B. Tiepolo 11, I-34131
Trieste, Italy; chiappini@ts.astro.it\\
$^5$Dipartimento di Astronomia, Universit\`a di Trieste, Via G.B. Tiepolo 11,
I-34131 Trieste, Italy; matteucci@ts.astro.it}

\begin{document}
\maketitle
\voffset=-1.0cm     

\begin{abstract} 

K-dwarfs have life-times older than the present age of the Galactic disc, and
are thus ideal stars to investigate the disc's chemical evolution.  We have
developed several photometric metallicity indicators for K dwarfs, based an a
sample of accurate spectroscopic metallicities for 34 disc and halo G and K
dwarfs. The photometric metallicities lead us to develop a metallicity index
for K dwarfs based only on their position in the colour absolute-magnitude
diagram. Metallicities have been determined for 431 single K dwarfs drawn from
the Hipparcos catalog, selecting the stars by absolute magnitude and removing
multiple systems. The sample is essentially a complete reckoning of the metal
content in nearby K dwarfs. We use stellar isochrones to mark the stars by
mass, and select a subset of 220 of the stars which is complete in a narrow
mass interval. We fit the data with a model of the chemical evolution of the
Solar cylinder. We find that only a modest cosmic scatter is required to fit
our age metallicity relation. The model assumes two main infall episodes for
the formation of the halo-thick disc and thin disc respectively. The new data
confirms that the solar neighbourhood formed on a long timescale of order 7
Gyr.
\end{abstract}


\section{Introduction}

A central issue in studies of the chemical evolution of the Galactic disc is to
resolve the so called ``G-dwarf problem'' (van den Bergh, 1962; Schmidt, 1963;
Pagel and Patchett, 1975). The problem is that the observed stellar metallicity
distribution shows far fewer metal deficient stars than the predictions of the
simplest, closed box models of the Galactic disc's chemical evolution.
Integrated light studies indicate that the G-dwarf problem is not restricted to
our own disc (Worthey et al, 1996) but is also found in other galaxies. There
are many ways that the evolutionary models can be modified to bring them into
consistency with the observations, such as pre-enrichment of the gas, a time
dependent Initial Mass Function or gas infall (for a review see e.g. Pagel,
1997)

G-dwarfs are sufficiently massive that some of them have begun to evolve away
from the main sequence, and these evolutionary corrections must be taken into
account when determining their space densities and metallicities. While these
problems are by no means intractable, it has been long recognized that K dwarfs
would make for a cleaner sample of the local metal abundance distribution,
because for these stars the evolutionary corrections are negligible.

K dwarfs are of course intrinsically fainter, and it has not been until
recently that accurate spectroscopic K dwarf abundance analyses have become
available, with which to calibrate photometric abundance indicators (Flynn and
Morell, 1997). As a result, it is now known that there is a K-dwarf problem
which is very similar to the G-dwarf problem (Flynn and Morell, 1997).  Studies
of M dwarfs indicate that the problem is present in these stars too (Mould
1976). Such stars are still not really ideal for measuring the local
metallicity distribution, because metallicities for M dwarfs are difficult to
determine in the optical (but appears possible using infrared $JHK$ photometry,
(see e.g. Stauffer and Hartmann, 1986; Leggett et al, 2000). The all sky surveys
presently underway in the infrared (Denis, 2MASS) may make such stars viable
metallicity tracers in the near future.

The G-dwarf metallicity distribution is already an extensively studied subject
(Pagel and Patchett, 1975; Sommer-Larsen, 1991; Wyse and Gilmore, 1995;
Rocha-Pinto and Maciel, 1996) and has been used by chemical evolution models to
constrain the time scale of the formation of the Galactic disc at the solar
neighbourhood. The G-dwarf metallicity distribution can be well fit by models
in which gas has been settling onto the disc over a protracted period, of some
billion years.  For example, Chiappini et al (1997), on the basis of the fit of
the G-dwarf metallicity distribution, have shown that the disc in the solar
neighbourhood was formed on a long time scale of 7 - 8 Gyr. This conclusion has
been later stressed also by other authors (e.g. Portinari et al, 1998; Prantzos
and Silk, 1998; Chang et al, 1999).

The release of the data from the European Space Agency's {\em Hipparcos} (ESA,
1997) satellite offers a great opportunity to determine the local metallicity
distribution of the disc from a complete sample of K dwarfs. There are three
clear improvements because of Hipparcos. Firstly, the parallaxes are so
accurate that the K dwarfs can be selected by absolute magnitude rather than
colour, which is a much better way of isolating stars in a particular mass
range. Secondly, the uniformity of the Hipparcos data allows us to construct
samples with well understood completeness limits. Thirdly, a large fraction of
the close binaries can be identified using Hipparcos and removed (since
photometric abundance indicators are calibrated for single stars). This last
effect turns out to be quite important.
  
In this paper we develop a number of simple photometric metallicity indicators
for K dwarfs, based on a spectroscopically determined sample of metallicities
by Flynn and Morell (1997). The metallicities for several hundred K dwarfs
drawn from the Hipparcos catalog have been measured via new observations
described here. This sample has been used to show that there is a very simple
relation between the $V$ band luminosity $M_V$ of K dwarfs at a given $B-V$
colour and metallicity (the work is fully described in a companion paper
(Kotoneva, Flynn and Jimenez, 2002, paper II). This simple relation is used in
this paper to determine metallicities for 431 single K dwarfs in a near
complete sample selected from the Hipparcos catalog. We examine a set of
isochrones and find a simple relation between the position of a star in the
Hipparcos colour magnitude diagram and mass. We are thus able to select a
sample of dwarf stars by the more physically relevant parameter of mass, rather
than by colour or absolute magnitude, as is usually the case. The metallicity
distribution obtained is then compared to the results of models of the chemical
evolution of the Solar neighbourhood.

The paper is organized as follows: In section 2 we describe the selection of
the K dwarf sample and in section 3 the observations and reductions of the
stars in broadband and intermediate band photometric systems. The details of
the various metallicity calibrations are presented in section 4. In section 5
we use isochrones to fit the masses of the stars, and select the stars in an
appropriate mass range which is near complete; we discuss the kinematics of the
stars and compute corrections for the metallicity distribution function from
the solar volume to the solar cylinder.  The metallicity distribution is
compared briefly with other determinations in the literature. We compare the
data to the predictions of a model of the chemical evolution of the Solar
neighbourhood due to Chiappini et al (2001) in section 6. We draw our
conclusions and summarize in section 7.

\section{The Sample}

\subsection{Sample selection}

Our sample of K dwarfs is drawn from the ESA Hipparcos catalog (ESA, 1997).
For the purposes of this paper we term K dwarfs to be stars of absolute
magnitude $M_V$ in the range $5.5 < M_V < 7.3$.  The choice of these absolute
magnitude limits is motivated as follows. The upper magnitude limit at $M_V =
5.5$ is chosen to avoid the effects of stellar evolution. Examination of
theoretical isochrones (see e.g. Jimenez, Flynn and Kotoneva, 1998) indicate
that the effects of stellar evolution on luminosity at $M_V = 5.5$ during the
disc lifetime amount to at most $0.1$ magnitude, typically much smaller. The
effects of stellar evolution in constructing the sample are thus small, and
negligible compared to the main source of error (which is Poisson sampling
statistics). The limit at $M_V = 5.5$ corresponds to a spectral type of about
G8. The lower absolute magnitude limit at $M_V = 7.3$ is the magnitude of the
reddest K dwarfs for which our photometric metallicity indicators can currently
be calibrated via spectroscopic observations (Flynn and Morell, 1997). The
limit at $M_V = 7.3$ corresponds a spectral type of about K5.

Since we are interested in obtaining a complete sample of K dwarfs in the Solar
neighbourhood, with which to construct the metallicity distribution function
(MDF), the stars were initially selected from the ``survey'' part of the
Hipparcos catalog, which is complete to an apparent visual magnitude given by
$V < 7.3 + 1.1\,{\mathrm sin}|b|$. Here $b$ is the Galactic latitude (the
apparent visual magnitude limit was made dependent on $b$ in order to avoid
observing excessive numbers of stars in the Galactic plane). Adopting this
apparent magnitude limit and the absolute magnitude range $5.5 < M_V < 7.3$
resulted in a sample of 209 stars. Our intention had been to get a sample of
order 500-750 stars. In order to increase our basic sample size, we increased
the magnitude limit by 0.9 mag (i.e. by taking an apparent magnitude limit of
$V < 7.3 + 1.1\,{\mathrm sin}|b| + 0.9$) we obtained 668 stars (including the
first 209 stars). This is the basic sample which we targeted for observations.

We need to understand the completeness of this basic sample, since it is drawn
from stars 0.9 magnitudes fainter than the magnitude limit of the complete part
of the Hipparcos catalog. Even 0.9 magnitudes beyond the completeness limit,
the sample turns out to be still satisfactorily close to complete for our
purposes.  The completeness level is approximately 94\%.  We determined this
quantity by using the Galactic structure model of Holmberg (2000). The model is
based on star count data and makes use of both the Hipparcos and Tycho data
(the latter is complete to much fainter limits than Hipparcos) to construct the
luminosity function of the local disc. Using the model, we find that for an
apparent magnitude limit of $V < 7.3 + 1.1\,{\mathrm sin}|b| + 0.9$ and
absolute magnitude limits of $5.5 < M_V < 7.3$, we would expect that some 710
stars should be in our basic sample, whereas there are actually 668 such stars
in the Hipparcos catalog. We conclude that even 0.9 magnitudes beyond the
``survey'' completeness limit, Hipparcos is substantially (94\%) complete in
the absolute magnitude range of interest.

Stars were observed by Hipparcos up to 4 magnitudes fainter than the faintest
stars in the ``survey''. The 0.9 magnitude extension beyond the survey limit
results in a sample which is still over 90\% complete. Beyond this 0.9 mag
limit, completeness drops rapidly, and the stars in Hipparcos become
increasingly dominated by objects included for a particular astrophysical
interest, such as having a low metallicity and/or high velocity. There will be
a very small excess of such stars in our extended sample, but the high
completeness (94\%) means that it is other uncertainties (Poisson statistics)
which dominate the construction of the metallicity distribution.

\begin{figure}
\epsfig{file=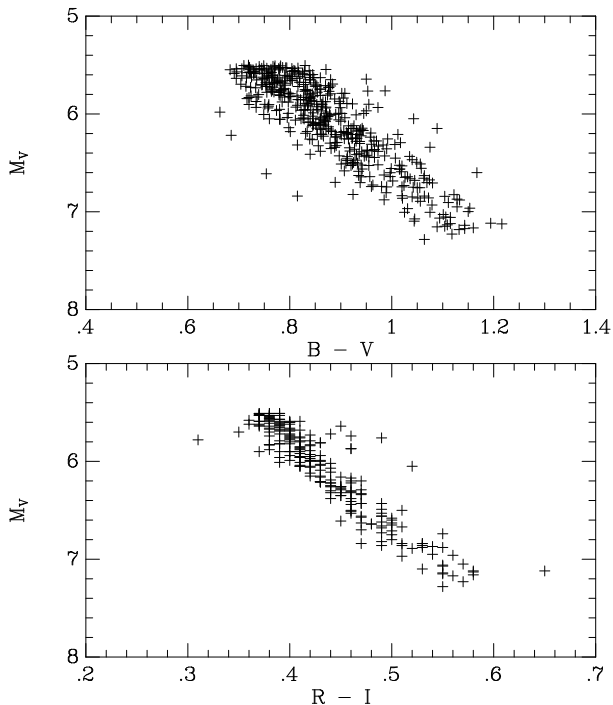,width=80mm}
\caption{The upper panel shows colour-magnitude diagram for 431 stars in our
sample for which $B - V$ colour is available, while the lower panel shows the
colour-magnitude diagram for the $R - I$ colour.  Note that the main sequence
is clearly thicker as a function of $B - V$ compared to $R - I$. This is a
direct consequence of the metallicity of the stars, which affects $B - V$ much
more than it affects $R-I$ (Flynn and Morell, 1997).}
\label{CMD}
\end{figure}

\subsection{Removal of multiple stars}

The photometric metallicity indicators described in section 4 were calibrated
using spectroscopically determined metallicities for single stars.  We examined
the effect that multiple stars would have on our metallicity indicators, by
making Monte-Carlo simulations in which we combined the fluxes of a range of
pairings of single K and M dwarfs selected at random from the Hipparcos
delineated main sequence, and computing the effect on the photometrically
determined metallicity. Metallicities for multiple stars were found to be as
much as $0.4$ dex lower than the true metallicity.  Cleaning the sample of
multiple stars thus turned out to be very important.

The Hipparcos catalog included a flag for ``probable multiple stars'', based
on the ``reliability of the double or multiple star solution''.  This flag was
used to eliminate all definite, possible and suspected multiple systems. This
reduced the sample from 668 to 449 stars or about 2/3 of the initial
sample. Despite this expedient, a small number of binaries seem to remain in
the sample (of order 10\%). This is evident from the positions of the suspected
multiples in the colour-magnitude and two colour $B-V$ versus $R-I$ diagrams.
This issue is studied in detail in a companion paper (Kotoneva, Flynn and
Jimenez, 2002, Paper II). About half of these extra suspected multiples could
be removed in constructing the final sample, so the final sample should have a
minor contamination by multiples of less than 5\%. Note that, by ``single
stars'', we mean only stars which have no companion bright enough to
significantly affect the metallicity measurement (i.e. within 5 magnitudes of
the brightness of the primary). After removing these suspected multiple stars
the final sample consists of 431 stars.

The colour-magnitude diagrams for our final sample of single stars are shown in
Fig \ref{CMD}.

\section{Observations and reductions}

Observations were made at Siding Spring Observatory (SSO) in Australia, between
$8^{\mathrm th}$ and $22^{\mathrm nd}$ of March 1999. We measured
Johnson-Cousins and Str{\"o}mgren colours for all available Southern hemisphere
stars of the basic sample using the SSO 24'' telescope.  We used the Motorized
Filter Box with all eight filters $U, B, V, R, I$ and Str{\"o}mgren $v, b$ and
$y$.  $UBVRI$ primary standards were selected from Landolt (1983a, 1983b, 1992)
and among E-region standard stars (Graham, 1982; Menzies et al.,
1989). Str{\"o}mgren primary standards were selected from the E-regions and
also from Gr{\o}nbech, Olsen, and Str{\"o}mgren (1976) and Crawford and Barnes
(1970). Hauck and Mermilliod (1998) stars were used as secondary standards for
the Str{\"o}mgren system and Flynn and Morell (1997) stars for both of the
intermediate band systems. We observed 218 stars and 55 of these more than
once. The standard error in the $R - I$ colour was $\approx 0.012$ mag, quite
accurate enough for our purposes. In the Str{\"o}mgren bands the photometric
error was slightly higher but still $\approx 0.02$ mag.

Further data were obtained at La Palma using the Swedish 60 cm telescope in
March 2000. An SBIG CCD was used to observe in the filters $V, R$ and $I$. 90
stars were observed, most of them more than once. The scatter for the La Palma
observations was bit higher than at SSO, the scatter in the $R-I$ band being
0.021 for a single observation. Since most of the stars were observed twice the
mean error came down to $\approx 0.015$ mag. 22 of the stars were observed both
in Australia and La Palma; the mean difference in $R - I$ was less than 0.01
mag and the scatter was 0.01 mag, so that no significant difference was found
between observations taken at the different observatories.

\begin{figure}
\epsfig{file=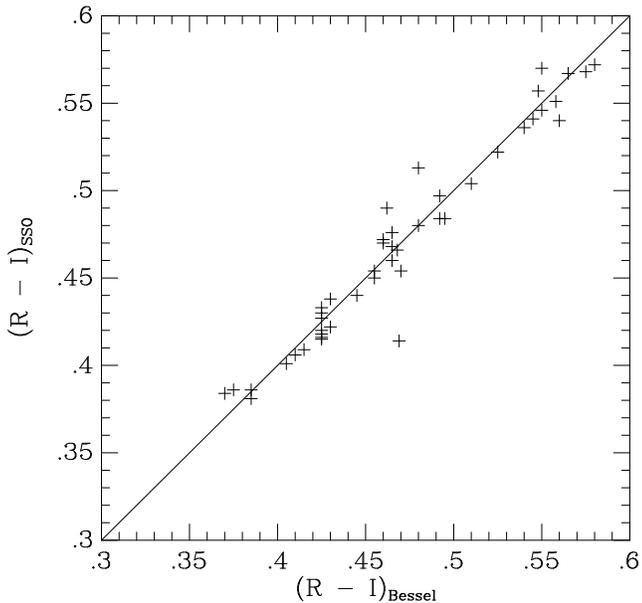,width=84mm}
\caption{Comparison of our $R - I$ colours obtained at SSO versus those from
Bessell's (1990) values from his observations of stars in the Gliese
catalog. The mean difference is less than 0.01 mag and the scatter is 0.013
mag.}\label{bes}
\end{figure}

We also obtained $R - I$ colours for 108 stars from Bessell (1990), who
observed all the stars in the Gliese catalog. These are found to be in
excellent agreement with our own SSO data with a mean difference in $R - I$ of
less than 0.01 mag with a scatter of 0.013 mag. A comparison of the two sources
of $R-I$ colour is shown in Fig \ref{bes}.

\section{Determination of Metallicities}

We have obtained metallicities for all 431 stars of the dataset. In the end, a
single method was used, although we discuss here four metallicity indicators
for the dwarfs.  Firstly, we used a relation based on Geneva and $R-I$
photometry due to Flynn and Morell (1997). Secondly, we have developed a very
similar method to Flynn and Morell's based on Str{\"o}mgren and $R-I$
colours. Thirdly, we have found a simple relation between the broadband $B-V$
and $V-I$ colours and metallicity. Fourthly, we have developed in a companion
paper (Kotoneva, Flynn and Jimenez, 2002) a metallicity indicator for K dwarfs
based on their absolute magnitude in the $V$ band relative to a fiducial solar
metallicity isochrone.  The first three methods have a similar error 0.2 dex in
$[{\rm Fe/H}]$, while the last method appears to be better, with an estimated
error of 0.1 dex.  This last method depends on the availability of accurate
parallaxes for the stars, and so is less generally applicable than those which
rely on photmetric colours alone. The last method is the one we adopt here for
measuring the K dwarf metallicities.

\subsection{Metallicities from Geneva photometry}

To obtain the metallicities for the K dwarfs for which a Geneva $b_1$ colour is
available, we used existing relations described in Flynn and Morell (1997). The
$b_1$ colours were obtained from Rufener (1989) for 245 stars and the $R - I$
colours come from our observations and/or from Bessell (1990). Both of the
colours were available for 149 stars for which the metallicities,
[Fe/H]$_{b_1}$, were computed using the relation:

\begin{equation}
{\rm [Fe/H]}_{b_1} = 8.248\times b_1 - 12.822\times (R - I) - 4.822.
\end{equation}

These metallicities have a typical error of 0.2 dex.

\subsection{Metallicities from Str{\"o}mgren photometry}

We have developed a new metallicity indicator for the K dwarfs for which
Str\"omgren colours are available. The calibration was obtained using the 34 G
and K dwarfs from Flynn and Morell (1997). For these stars accurate,
spectroscopically determined metallicities, [Fe/H]$_{\mathrm spec}$ and
effective temperatures have been determined with errors of 0.05 dex and
$\approx 100$ K, respectively.

We searched for a relation between $R - I$ colour, Str\"omgren colours and
spectroscopical abundance, $[{\rm Fe/H}]_{\mathrm spec}$. The best fitting
Str\"omgren colour was found to be $m_1$, with a small dependence on $c_1$. The
$b - y$ colour was found to be poorly correlated with metallicity as one might
expect. The relation we found is:

\begin{eqnarray}
{\rm [ Fe/H]}_{m_1} & =& 6.076\times {\rm m_1} -3.811 \times{\rm c_1} +
9.4303\times{\rm c_1}^2\\ & & -12.822\times {\rm (R - I)} + 3.226 \nonumber.
\end{eqnarray}

\begin{figure}
\epsfig{file=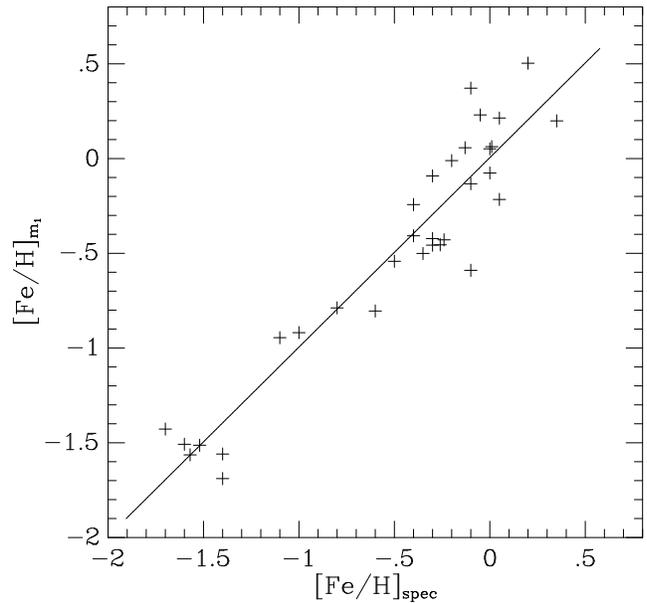,width=84mm}
\caption{Relation between the metallicities using Str{\"o}mgren $m_1$ colour
and spectroscopically determined metallicities.  The scatter around the 1:1
line is $\approx 0.20$ dex.}\label{tran}
\end{figure}

Fig \ref{tran} shows the relation between the spectroscopically determined
metallicity and the Str\"omgren and $R-I$ based metallicity. The scatter around
the one-to-one relation is $\approx 0.20$ dex.

For 142 stars both the Str{\"o}mgren $m_1$ and $c_1$ colours and $R - I$ colour
were available and the metallicities $[{\rm Fe/H}]_{m_1}$ were calculated. The
dependence on $c_1$ is quite weak, so that while for 54 stars no $c_1$ colour
was available, for these stars we adopted a typical value of $c_1 =
0.29$. Adopting this mean value leads to a very small increase in the
metallicity error of $\approx$ 0.02 dex.

\subsubsection{A check using the Hyades}

\begin{figure}
\epsfig{file=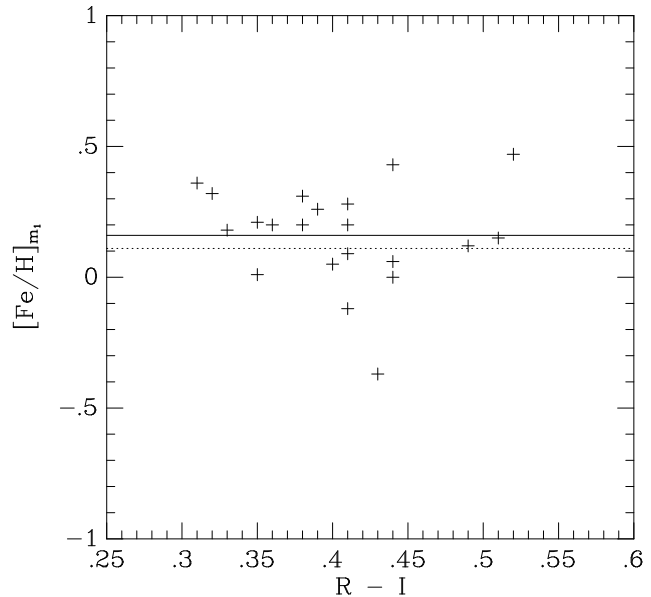,width=84mm}
\caption{The metallicities $[{\rm Fe/H}]_{m_1}$ for Hyades G and K dwarfs as a
function of $R - I$. The dotted line represents the mean metallicity value from
the literature $[{\rm Fe/H}] = 0.11$ (Taylor, 1994). Our mean value $[{\rm
Fe/H}] = 0.16$ is shown as a solid line. }\label{hyad}
\end{figure}

A check of the Str{\"o}mgren calibration was made using G and K dwarfs in the
Hyades cluster (Reid, 1993; Flynn and Morell, 1997), using $VRI$ photometry and
Str{\"o}mgren $m_1$ colour from the literature (Hauck and Mermilliod, 1998). A
list of the Hyades G and K dwarfs, their magnitudes, $B -V$, $R - I$ and $m_1$
colours and the metallicities calculated using eqn. (2) are shown in Table
\ref{tab1}. Taylor (1994) estimated the mean metallicity for the Hyades to be
$[{\rm Fe/H}] = 0.11 \pm 0.01$. Our mean value is $[{\rm Fe/H}] = 0.16 \pm
0.03$. Fig \ref{hyad} shows the relation between the derived metallicity and
the $R - I$ colour, and is found to be independent of colour, indicating that
there are no residual temperature effects in the metallicity indicator.

\begin{table}
\begin{center}
\caption{Broadband and Str\"omgren $m_1$ data for G and K dwarfs in the
Hyades}
\begin{tabular}{r|c|c|c|r}
\hline
HD no.  & $V $ & $R - I$ & $m_1$ & $[{\rm Fe/H}]_{m_1}$ \\
\hline
 26756  &  8.46 &  0.35 &  0.25 &  0.01\\
 26767  &  8.04 &  0.31 &  0.21 &  0.36\\
 27771  &  9.09 &  0.39 &  0.38 &  0.26\\
 28099  &  8.10 &  0.32 &  0.23 &  0.01\\
 28258  &  9.02 &  0.43 &  0.36 &$-$0.37\\
 28805  &  8.66 &  0.35 &  0.29 &  0.21\\
 28878  &  9.38 &  0.41 &  0.42 &  0.20\\
 28977  &  9.65 &  0.44 &  0.45 &  0.00\\
 29159  &  9.37 &  0.41 &  0.40 &  0.09\\
 30246  &  8.31 &  0.33 &  0.24 &  0.18\\
 30505  &  8.98 &  0.38 &  0.37 &  0.31\\
 32347  &  8.98 &  0.36 &  0.31 &  0.20\\
284253  &  9.14 &  0.38 &  0.35 &  0.20\\
284787  &  9.05 &  0.40 &  0.37 &  0.05\\
285252  &  9.00 &  0.41 &  0.43 &  0.28\\
285690  &  9.56 &  0.44 &  0.52 &  0.43\\
285742  & 10.26 &  0.49 &  0.59 &  0.12\\
285773  &  8.94 &  0.41 &  0.36 &$-$0.12\\
285830  &  9.47 &  0.44 &  0.45 &  0.06 \\
286789 &  10.44 &  0.52 &  0.70 &  0.47\\
286929 &  10.01 &  0.51 &  0.63 &  0.15\\
\hline
\end{tabular}
\label{tab1}
\end{center}
\end{table} 

\subsection{Comparison of the Str\"omgren and Geneva photometric 
metallicity calibrations}

\begin{figure}
\epsfig{file=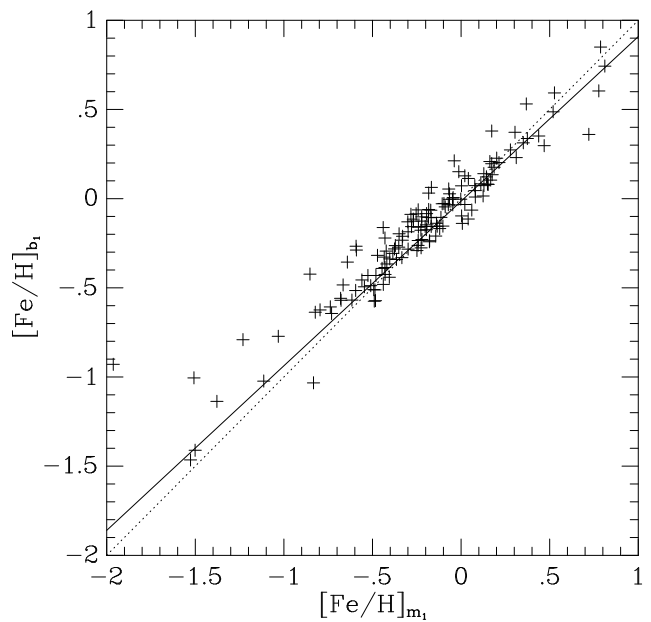,width=84mm}
\caption{The relation between the metallicities based on the Str{\"o}mgren
$m_1$ and Geneva $b_1$ colours. The solid line shows a least square fit to the
data (where the errors in both colours are taken into account). The dotted line
shows the 1:1 relation.}\label{mrela}
\end{figure}

In Fig \ref{mrela} the relation between the metallicities calculated using the
Geneva colour $b_1$ and the Str\"omgren $m_1$ colour is shown. The separate
metallicities are in good agreement with each other: the scatter around the
fitted line is only $0.15$ dex, significantly less than the error of the
individual metallicity estimates of 0.2 dex. This is probably because the $b_1$
and $m_1$ colours are correlated, since they both measure similar regions in
the blue at approximately 4000 {\AA}. The metallicities are found to be in
excellent agreement for $[{\rm Fe/H}] > -0.7$, where most of the stars lie,
differing by less than 0.01 dex. For the stars with metallicities $[{\rm Fe/H}]
< -0.7$, the scatter appears to increase and there may be some systematic shift
between the systems. In terms of the K dwarf problem, a possible systematic
error at such low metallicity is not very significant.  More low metallicity
stars with good spectroscopic metallicity determinations would be of interest
for testing the abundance indicators further.

\subsection{Metallicities from broadband photometry}

\begin{figure}
\epsfig{file=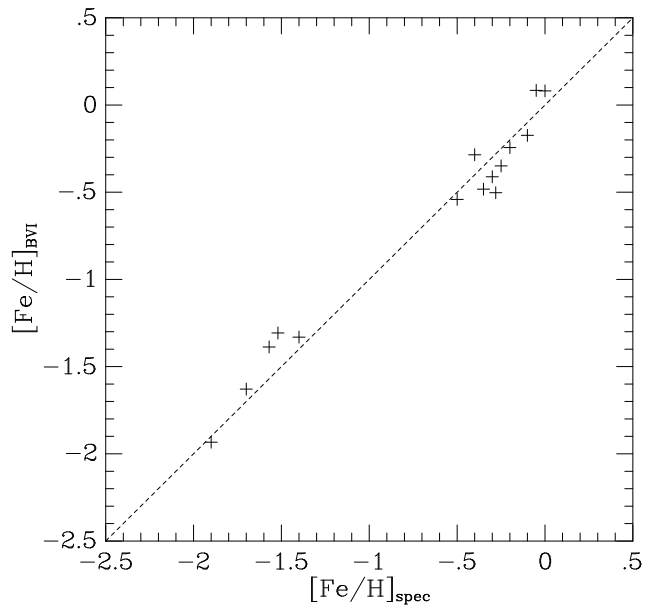,width=84mm}
\caption{Relation between the metallicities using the broadband colours $B-V$ and
$V-I$ for K dwarfs in the absolute magnitude range $5.5 < M_V < 7.3$. The
scatter around the 1:1 line is $\approx 0.1$ dex.}\label{fehbvi}
\end{figure}

We have also developed a new metallicity indicator for K dwarfs using broadband
$BVI$ photometry. For this purpose we only choose calibration K dwarfs in the
absolute magnitude range $5.5 < M_V < 7.3$. The calibration stars were selected
from the Flynn and Morell (1997) sample, with $B-V$ and $V-I$ data being
collated from Bessell (1990). The calibrating sample is shown in Table
\ref{bvi}. We recovered a simple relation between $B-V$, $V-I$ colour and
metallicity as follows:

\begin{equation}
[\mathrm Fe/H]_{\mathrm BVI} = 8.54\times (B-V) - 7.73 \times (V-I) - 0.55
\end{equation}

where we denote the metallicity index based on these colours by
[Fe/H]$_{\mathrm BVI}$.  A comparison of the spectroscopic
metallicities and this broadband photometric metallicity indicator is
shown in Fig \ref{fehbvi}. The scatter in the fit is $\approx 0.1$
dex. Note that this simple relation should not be used for stars with
absolute magnitude $M_V < 5.5$, because the effects of stellar
evolution begin to affect the colours in addition to the metallicity.

\begin{table}
\begin{center}
\caption{K dwarfs used to calibrate a photmetric metallicity indicator based on
$B-V$ and $V-I$ colours. Columns are the HD number, colours, absolute magnitude
$M_V$, the spectroscopically measured metallicity [Fe/H]$_{\mathrm spec}$ and
the photometric metallicity [Fe/H]$_{\mathrm BVI}$ (Eqn. 3)}
\begin{tabular}{rrrrrr}
\hline
HD no. &$B-V$  & $V-I$ & $M_V$ &[Fe/H]$_{\mathrm spec}$ & [Fe/H]$_{\mathrm BVI}$ \\ 
\hline
  4628 & 0.890 & 0.949 & 6.376 &$ -0.40$ &$ -0.29 $ \\
 10700 & 0.727 & 0.802 & 5.680 &$ -0.50$ &$ -0.54 $ \\
 13445 & 0.812 & 0.891 & 5.930 &$ -0.28$ &$ -0.50 $ \\
 25329 & 0.863 & 1.093 & 7.178 &$ -1.70$ &$ -1.63 $ \\
 26965 & 0.820 & 0.888 & 5.916 &$ -0.30$ &$ -0.41 $ \\
 64090 & 0.621 & 0.865 & 6.008 &$ -1.90$ &$ -1.93 $ \\
 72673 & 0.780 & 0.853 & 5.953 &$ -0.35$ &$ -0.48 $ \\
100623 & 0.811 & 0.870 & 6.063 &$ -0.25$ &$ -0.35 $ \\
103095 & 0.754 & 0.934 & 6.611 &$ -1.40$ &$ -1.33 $ \\
134439 & 0.770 & 0.959 & 6.736 &$ -1.57$ &$ -1.39 $ \\
134440 & 0.850 & 1.037 & 7.077 &$ -1.52$ &$ -1.31 $ \\
149661 & 0.827 & 0.832 & 5.819 &$  0.00$ &$  0.08 $ \\
192310 & 0.878 & 0.888 & 6.002 &$ -0.05$ &$  0.08 $ \\
209100 & 1.056 & 1.118 & 6.893 &$ -0.10$ &$ -0.17 $ \\
216803 & 1.094 & 1.169 & 7.065 &$ -0.20$ &$ -0.24 $ \\
\hline
\end{tabular}
\label{bvi}
\end{center}
\end{table}

The metallicity index is quite accurate, and is based on broadband colours
only. As a consequence it is probably more useful than the indices based on the
narrower band Str\"omgren and Geneva colours developed in the previous
sections. A simple test of the calibration could be obtained by selecting low
reddening open and globular clusters for which sufficiently deep main sequence
photometry in $B, V$ and $I$ is available, since metallicities for these are
known from spectroscopic studies.

By applying this type of metallicity indicator to large photometric surveys
(such as Sloan) it would be in principle possible to estimate the metallicity
distribution of K dwarfs at different Galactocentric distances in the Milky
Way's disk and even in the bulge region. This would represent an powerful
constraint for chemical evolution models. Moreover this metallicity indicator
could also be used to study local group galaxies by utilizing deep two color
data obtained with Hubble Space Telescope.

\subsection{Metallicities using K dwarf luminosity and colour}\label{metscatt}

We show in a companion paper (Kotoneva, Flynn and Jimenez, 2002) that stellar
luminosity on the main sequence correlates very well with metallicity at a
given colour. This was shown by measuring the displacement of stars $\Delta
M_V$ of K dwarfs from a fiducial isochrone in the $M_V$ versus $B-V$ plane
(i.e. relative to the fiducial line at that colour). The fiducial isochrone
comes from Jimenez, Flynn and Kotoneva (1998), has an age of 11 Gyr, solar
metallicity, and was found empirically to be a good fit to solar metallicity K
dwarfs.  The metallicity determined by this method, [Fe/H]$_{\rm KF}$, is given
by:

\begin{equation}
{\rm [Fe/H]_{\mathrm KF}} = 1.185 \times \Delta M_V + 0.054.
\end{equation}

This new metallicity indicator was developed on the basis of the photometric
derived metallicities described above. When we checked that [Fe/H]$_{\mathrm
KF}$ is consistent with the sample of K dwarfs with spectroscopically
determined metallicities (and in which it is ultimately based, since the
photometric metallicities have been calibrated from these same dwarfs) the
indicator turned out to be considerably more accurate than anticipated. The
scatter between the [Fe/H]$_{\mathrm KF}$ and the spectroscopically measured
metallicities was $\approx 0.08$ dex, i.e. little more than the intrinsic error
in the spectroscopic metallicities (0.05 dex). Although the spectroscopic
sample is rather small, we consider that metallicities derived using this
technique are greatly superior to photometrically derived metallicities. Note
that the technique relies on accurate parallaxes (absolute magnitudes) being
available, and is at present applicable only to nearby stars (or stars for
which an independent distance indicator is available). Alternatively, the
photometric techniques developed in this paper, although of lower precision,
can be used on stars for which colours only are available.

For all the stars in the sample we show the metallicity [Fe/H]$_{\mathrm KF}$
in column 12 of Table \ref{database}. The values are based on the Hipparcos
$B-V$ and $M_V$. The error in [Fe/H] is dominated by the colour error, which
for the sample stars is typically 0.025 mag, leading to a typical error in the
metallicities of 0.1 dex.

\section{K dwarfs in the Hipparcos catalog}

\subsection{The data}

Part of the full dataset of 431 stars is shown in Table \ref{database}. The
table shows the Hipparcos Input catalog (HIP) and HD numbers, visual magnitude
$V$ and absolute magnitude $M_V$, as computed from the Hipparcos parallax. The
next five columns are the $B - V$ colour, the mean values of $R - I$ and $m_1$
and $c_1$ Str{\"o}mgren colours and $b_1$ from the Geneva catalog. The
Str{\"o}mgren based metallicity, ${\rm [Fe/H]}_{m_1}$, Geneva based metallicity
${\rm [Fe/H]}_{b_1}$ and the luminosity based metallicity ${\rm [Fe/H]}_{KF}$
follow. The three last columns show the $U, V$ and $W$ velocities in \kms. The
full dataset is available at the Strasbourg Data Center or from the authors.

\begin{table*} 
\begin{center}
\caption{Part of the dataset showing identifications (HIP and HD numbers),
visual $V$ magnitude, absolute magnitude $M_V$, broadband colours $B-V$ and
$R-I$, Str\"omgren colours $m_1$ and $c_1$, the Geneva colour $b_1$, abundance
estimates from various methods (see section 5) and the space velocities $U, V$
and $W$. The full table of 431 stars is available at the Strasbourg Data Center
or from the authors.}
\begin{tabular}{r|c|c|c|c|l|r|r|r|r|r|r|r|r|r}
\hline
HIP & HD & $V $ & $M_V$ & $B - V $ & $R - I$ & $m_1$ & $c_1$ & $b_1$ 
& $[{\rm Fe/H}]$ & $[{\rm Fe/H}]$ & $[{\rm Fe/H}] $& $U$ & $V$ & $W$ \\ 
     &        &      &      &     &    &     &    &     & $m_1$ & $b_1$ & KF &                &   &  \\
\hline
 1031 &    870 & 7.22 &  5.68 &  0.77  &  --  &  0.29 &  0.28 & 1.17 &  --     & --      & $-$0.27 &  --    &    -- & --    \\
 1085 &    924 & 9.05 &  6.44 &  0.91  &  --  &  0.42 &  0.28 &  --  &  --     & --      & $-$0.35 &  --    &    -- & --    \\
 1837 &   1910 & 8.74 &  7.01 &  1.08  &  --  &   --  &   --  & 1.40 &  --     & --      & $-$0.26 &  --    &    -- & --    \\
 1936 &   2025 & 7.92 &  6.64 &  0.94  & 0.48 &  0.47 &  0.23 & 1.28 & $-0.43$ & $-0.39$ & $-$0.45 &  $-$41 &  $-$7 & $-$1  \\
 2194 &   2404 & 9.02 &  5.70 &  0.75  &   -- &  0.24 &  0.26 & 1.13 &  --     & --      & $-$0.44 &  --    &    -- & --    \\
 2736 &   3167 & 8.97 &  5.70 &  0.83  &   -- &   --  &   --  &  --  &  --     & --      &    0.08 &  --    &    -- & --    \\
 2742 &   3141 & 8.02 &  5.71 &  0.87  &   -- &  0.41 &  0.31 &  --  &  --     & --      &    0.29 &     32 &    25 &  0    \\
 2743 &   3222 & 8.55 &  6.21 &  0.85  &   -- &  0.37 &  0.29 & 1.24 &  --     & --      & $-$0.41 &  $-$64 &$-$102 & 44    \\
 3028 &   3569 & 9.21 &  6.11 &  0.85  &   -- &    -- &    -- &  --  &  --     & --      & $-$0.30 &  --    &    -- & --    \\
 3206 &   3765 & 7.36 &  6.17 &  0.94  &   -- &  0.49 &  0.30 & 1.30 &  --     & --      &    0.10 &     29 & $-$67 & $-$20 \\
 3535 &   4256 & 8.03 &  6.32 &  0.98  & 0.46 &    -- &   --  & 1.35 &  --     & 0.42    &    0.10 &     42 & $-$23 & $-$27 \\
\hline
\end{tabular}\label{database}
\end{center}
\end{table*}

\subsection{Raw metallicity distribution function for K dwarfs}

\begin{figure}
\epsfig{file=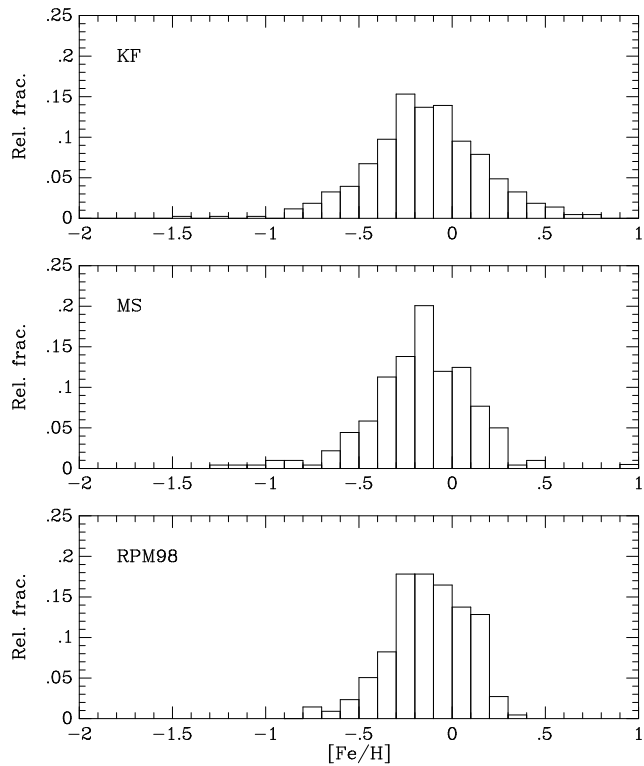,width=84mm}
\caption{Normalised metallicity distribution functions (MDF) for K dwarf sampes
in the literature and for this study. Lower panel : Rocha-Pinto and Maciel
(1998b).  Middle panel: Marsakov and Shevelev (1988). Upper panel: the 431 stars
in our basic sample, before any corrections.  The three samples all show a
similar spread in abundance and a peak at $[{\rm Fe/H}] \approx -0.2$. The
present sample (upper panel) clearly shows a longer tail of super-solar
metallicity stars; this is a consequence of the selection of the stars by
absolute magnitude. In section \ref{Kdwarfmasses} we develop a selection by
stellar mass which substantially reduces the number of stars at super-solar
metallicities.}
\label{MDFcomp}
\end{figure}

The normalised metallicity distribution function (MDF) for our basic sample of
431 K dwarfs is shown in the upper panel of Fig \ref{MDFcomp}, and is based
on the [Fe/H]$_{\mathrm KF}$ metallicities.  There are two studies of the
metallicity distribution of K dwarfs in the literature with which we can
compare our results. We compare the raw MDF, because this most closely matches
the procedures which have been used to select K dwarf samples in the past
(rather than comparing to the more carefully selected sample of K dwarfs
described in the next section).

Fig \ref{MDFcomp} also shows the metallicity distribution functions
obtained by Marsakov and Shevelev (1988) and Rocha-Pinto and Maciel
(1998b, RPM98). The Marsakov and Shevelev sample is an inhomogeneous
compilation of metallicities from the literature. The Rocha-Pinto and
Maciel stars were selected from the Third Gliese catalogue (Gliese et
al, 1991), for which photometric data could be found from the
literature. The present sample is a near complete sample of the solar
neighbourhood based on Hipparcos parallaxes and new photometric
observations. The samples have rather different pedigrees, but they
are all similar to the metallicity distribution for G dwarfs, being
peaked at $\simeq -0.2$ dex.

The clear result in Fig \ref{MDFcomp} is that the metallicity distribution we
obtain here is broader than the other samples. We believe this can be
attributed to the selection of our sample by absolute magnitude rather than by
spectral type.  We will argue this view in section \ref{Kdwarfmasses}, and use
what we regard as a superior procedure of selecting the stars by mass, rather
than spectral type, colour or luminosity, as has been done in the past.

\begin{figure}
\epsfig{file=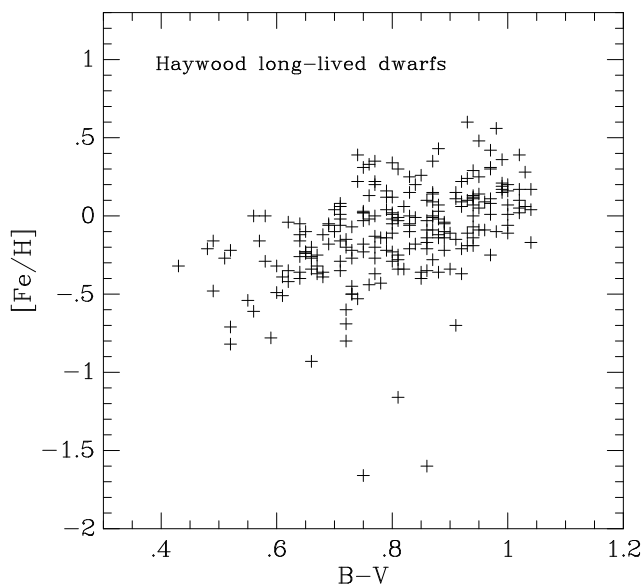,width=84mm}
\caption{Colour versus metallicity for the Haywood (2001) sample of long-lived
dwarfs. There is a systematic change on the metallicity distribution with
stellar colour, with the super-solar metallicity stars dominating amongst the
cooler spectral types.}
\label{haywood}
\end{figure}

Haywood (2001) has shown that the metallicity distribution of nearby G dwarfs
for all samples in the literature peak at ${\rm [Fe/H]} \approx -0.2$, as does
the present K-dwarf sample and the RPM98 sample. Haywood (2001) however argues
that the peak should be at ${\rm [Fe/H]} = 0.0$, based on samples selected by
colour and not spectral type, as has been done in the past. Our sample is
selected by absolute magnitude rather than colour, since this more reliably
selects the stars by mass, yet we still find the peak at its traditional
location. A comparison of 101 stars in common between Haywood's and our sample
does, indeed, show an offset between our stars and his of 0.2 dex. Furthermore,
a comparison with 40 stars in common with Rocha-Pinto and Maciel (1998b) shows
an offset of 0.15 dex (in the same sense as the Haywood comparison). These two
comparisons are not independent because the metallicities derived for the stars
are partially based on Str\"omgren photometry in both cases. Closer comparison
with the Haywood sample shows that the offset is a linear function of the
absolute magnitude of the stars, rising from $\approx 0.0$ at $M_V = 5.5$ to
$\approx 0.4$ at $M_V = 7.3$. Fig \ref{haywood} shows the Haywood stars (the
``long-lived dwarfs'') in $B-V$ colour versus [Fe/H]. The super-solar
metallicity stars dominate preferentially among the cooler dwarfs, which
suggests that there may be a systematic error in the metallicities as a
function of colour (i.e. effective temperature) or some selection
effect. Subsequent to noting this trend, we found that it has already been
commented upon by Reid (2002). We note that restricting the Haywood sample to
$B-V < 0.7$ would put the peak in the metallicity distribution at its
traditional location (as we find here) at ${\rm [Fe/H]} \approx -0.2$. We will
later select our K dwarf sample by mass (section \ref{Kdwarfmasses}).  The
metallicity distribution is not significantly altered if we divide the K dwarfs
into subsamples by mass (see Fig \ref{fehmass}) within our adopted mass limits.

\subsection{K dwarf kinematics}

The basic sample consists of 431 K dwarf stars with metallicity estimates
accurate to $\approx 0.1$ dex. For 212 of the stars, radial velocities were
found in the literature and space velocities $U$, $V$ and $W$
computed. Unfortunately, velocities are not available for all the stars; thus
there is likely to be a small bias toward higher velocity stars in the
literature sources, since high proper motion and metal weak stars are
preferentially included in radial velocity programs. However, since about half
the sample does have radial velocities, and the sample is furthermore dominated
by thick disc and disc stars which have much smaller proper motions than halo
stars, the bias in the measured velocity dispersions is likely to be quite
small. A useful extension to this work would be to obtain velocities for all
the stars.

\begin{figure}
\epsfig{file=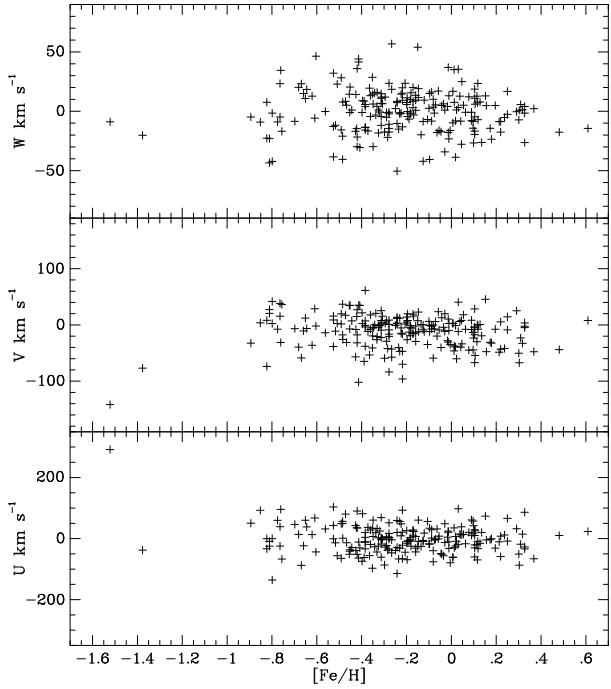,width=80mm}
\caption{Space motions $U$, $V$ and $W$ of the K-dwarfs as a function of the
metallicity [Fe/H]$_{\mathrm KF}$. Note that the velocity scale is different in
each panel.}\label{nop}
\end{figure}

\begin{figure}
\epsfig{file=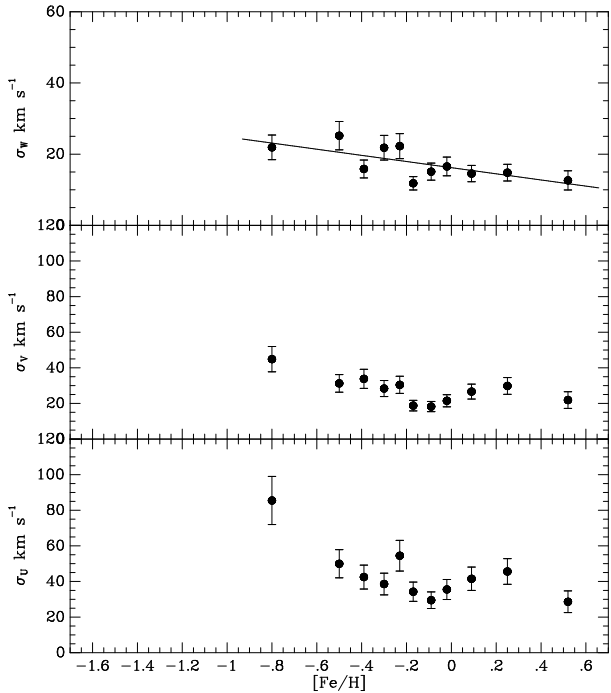,width=80mm}
\caption{Velocity dispersions $\sigma_U$, $\sigma_V$ and $\sigma_W$ of the K
dwarfs versus the metallicity [Fe/H]$_{\mathrm KF}$. The vertical velocity
dispersion $\sigma_W$ has been fit linearly with metallicity (upper panel) and
is used to correct the raw metallicity distribution function for the effect of
the scale heights of the stars. }\label{sigm}
\end{figure}

The velocities are shown as a function of metallicity in Fig \ref{nop}. The
velocity dispersions $\sigma_U, \sigma_V$ and $\sigma_W$ are shown as a
function of metallicity in Fig \ref{sigm}.  The figures show the expected
features of the solar neighbourhood: the thin disc with vertical velocity
dispersion of about 20 \kms for metallicities above approximately [Fe/H]
$=-0.5$, the kinematically hotter thick disc in the range $-1 \la $ [Fe/H] $\la
-0.5$.  There are very few stars with halo-like kinematics (high velocity
dispersion and high asymmetric drift). This is as one would expect: for a local
sample consisting initially of 431 K dwarfs, one would only expect of order 1
halo star, since the local disc:halo normalization is approximately 500:1
(Gould, Flynn and Bahcall, 1998). The apparent low metallicity stars are more
likely the result of scatter from stars in the thick disc, since there is an
observational scatter in the metallicities of 0.1 dex.

\subsection{Scale height/velocity correction}\label{shvel}

Models of the chemical evolution of the local disc predict the metallicity
distribution in a column through the disc, whereas the sample K dwarfs are
drawn from a roughly spherical region centered on the Sun. The local sample is
therefore biased toward stars of lower velocity dispersion since they will
spend more time close to the Galactic mid-plane than older, faster moving
stars.

We correct for this by computing the velocity dispersion of the stars as a
function of metallicity, and computing from this their vertical scale height
using a realistic mass model of the Galactic disc (following Sommer-Larsen,
1991).

\begin{table}
\begin{center}
\caption{Scale height correction for the metallicity distribution. The first
column shows the metallicity ranges of the bins, the second column shows the
observed velocity dispersion of the stars in the bin, $\sigma_W$(obs), the
third column the fitted vertical velocity dispersion, $\sigma_W$(fit) (i.e. to
the linear relation shown in the upper panel of Fig \ref{sigm}). The final
column shows the ratio $f$ of the column density to the local density of the K
dwarfs computed in a model of the Galactic disc (normalised so that $f = 1.0$
at [Fe/H] $=-0.02$). The total mass density of the K dwarfs in the model is
0.0043 $\Msun pc^{-3}$ and the total column density ($|z| < 2000.0$ pc) is 3.5
$\Msun pc^{-2}$.}
\begin{tabular}{r|c|c|c}
\hline
[Fe/H]  & $\sigma_W$(obs)    &  $\sigma_W$(fit)    &  $f$  \\
        & ${\rm km s^{-1}}$  &  ${\rm km s^{-1}}$  &       \\
\hline
$-0.80$ & 21.9 & 23.1 & 1.99 \\
$-0.50$ & 25.2 & 20.5 & 1.57 \\
$-0.39$ & 15.9 & 19.6 & 1.43 \\
$-0.30$ & 21.8 & 18.8 & 1.32 \\
$-0.23$ & 22.3 & 18.2 & 1.23 \\
$-0.17$ & 11.8 & 17.7 & 1.16 \\
$-0.09$ & 15.1 & 17.0 & 1.07 \\
$-0.02$ & 16.6 & 16.4 & 1.00 \\
 $0.09$ & 14.6 & 15.4 & 0.89 \\
 $0.25$ & 14.8 & 14.0 & 0.74 \\
 $0.52$ & 12.6 & 11.7 & 0.51 \\
\hline
\end{tabular}
\label{ftab}
\end{center}
\end{table} 

The vertical velocity dispersions are tabulated in Table \ref{ftab} and shown
in the upper panel of Fig \ref{sigm}. In order to smooth out noise due to
the small sample size, we have fit the vertical velocity dispersion linearly as
a function of metallicity (shown as a solid line in the upper panel of Fig
\ref{sigm}. We use the fit values of the velocity dispersion in what follows to
determine the correction of the MDF for the scale height of the stars.

The volume density of matter in main sequence dwarfs in the absolute magnitude
range $5 < M_V < 8$ is 0.0074 \Msun pc$^{-3}$ (Holmberg and Flynn, 2000, Table
1). The part of this which is represented by our sample K dwarfs ($5.5 < M_V <
7.3$) is 0.0043 \Msun pc$^{-3}$ (Holmberg, 2001, private communication). We
have determined the vertical velocity dispersion for the K dwarfs as a function
of metallicity, by sorting the stars by metallicity and dividing them into 11
equal star number bins (so that each bin represents a local volume density of
$0.0043/11 = 0.00039$ \Msun pc$^{-3}$).  From this local density and the
velocity dispersion of each bin, we then compute the total column density
represented by each bin by integrating self-consistently the Poisson-Boltzmann
equations in a model of the local Galactic disc (Holmberg and Flynn, 2000).  A
correction factor $f$, which is the ratio of the column density to the local
density for each bin, normalised so that $f=1$ in the bin closest to the solar
abundance (i.e. at [Fe/H] $= -0.02$), is computed and is shown in Table
\ref{ftab}. 


\subsection{Chromospheric activity corrections}

In this section we investigate the effects on the metallicity distribution of G
or K dwarfs of chromospheric activity.

Due to the Wilson-Bappu effect (Wilson and Bappu, 1957; Wilson, 1976),
photometrically derived metallicities differ somewhat from spectroscopic ones
(RPM98, Rocha-Pinto and Maciel 1998a and references therein). The Wilson-Bappu
effect in active chromospheres causes an emission line in the center of the
stellar absorption lines, which leads to too low photometrically determined
[Fe/H] metallicities.  If a sample includes many active stars, the metallicity
distribution would be biased towards metal poor stars.

The effect on the present sample turns out to be quite small.  Firstly,
chromospheric activity is stronger among binary stars. In the present sample
the binaries have been very effectively removed due to the high quality
Hipparcos data for all the stars. The sample is therefore similar to RPM98, in
which the binaries were also removed. Following RPM98, we can expect that about
30\% of the stars the sample are chromosphericaly active after removing the
binaries. The metallicity of these stars will be systematically incorrect.

RPM98 have defined a correction to the metallicity distribution, for
metallicities determined by Str\"omgren photometry, for G and K dwarfs.  A
similar analysis as in RPM98 has been carried out for Geneva photometry (a
medium band filter system quite similar to the Str\"omgren system), using those
stars in our sample for which accurate spectroscopic metallicities and
measurements of the emission line width $R_{HK}$ are available (Rocha-Pinto,
private communication).  Although this yielded only 8 stars, the same trend was
found for Geneva based metallicities as was found by RPM98 for Str{\"o}mgren
based metallicities. Detailed computations show that the effect of this
correction is to shift the mean of the metallicity histogram by $\approx +0.05$
dex, which can be easily understood since $\approx 30$\% of the stars are
thought to be chromospherically active and the mean correction to ${\rm
[Fe/H]}$ for such stars is 0.15 dex (RPM98, their section 4).

In the final sample of K dwarfs, we used metallicities based on the position of
the stars in the Hipparcos colour magnitude ($B-V$ versus $M_V$) diagram,
rather than the purely colour based (intermediate band) metallicities above.
Campbell (1984) has shown that the $B-V$ colours of lower main sequence stars
in the Hyades are affected by stellar activity, with typical changes in the
$B-V$ colour of order $\pm 0.02$ mag; this would change the measured
metallicities for the active stars by approximately $\pm 0.1$
dex. Unfortunately, without a detailed study, we cannot be sure if the effect
would also yield a systematic error in the metallicities.

In light of the above, we have decided not to make a chromospheric activity
correction to the sample because we cannot quantify its effect well enough.
The correction is likely to be small ($\approx 0.05$ dex), and probably shifts
the position of the peak of the metallicity distribution rather than changing
its shape. A major result of the paper, that the K dwarfs have a similar
metallicity distribution to the G dwarfs, is therefore robust despite the
uncertainties surrounding what chromospheric corrections should be applied, if
any.

\subsection{K dwarf masses}\label{Kdwarfmasses}

In the past, the metallicity distribution function (MDF) of the local disc has
been determined for stars of a particular spectral type or in some colour
range. However, in the context of modelling the chemical evolution of the
Galaxy and fitting it to the observed MDF, it is the range of stellar masses
represented in the MDF which is of most interest. Selection of the stars within
a particular range of masses permits a more direct comparison between the
observations and theory. We describe in this section the use of isochrones to
estimate masses for the K dwarfs.

\begin{figure}
\epsfig{file=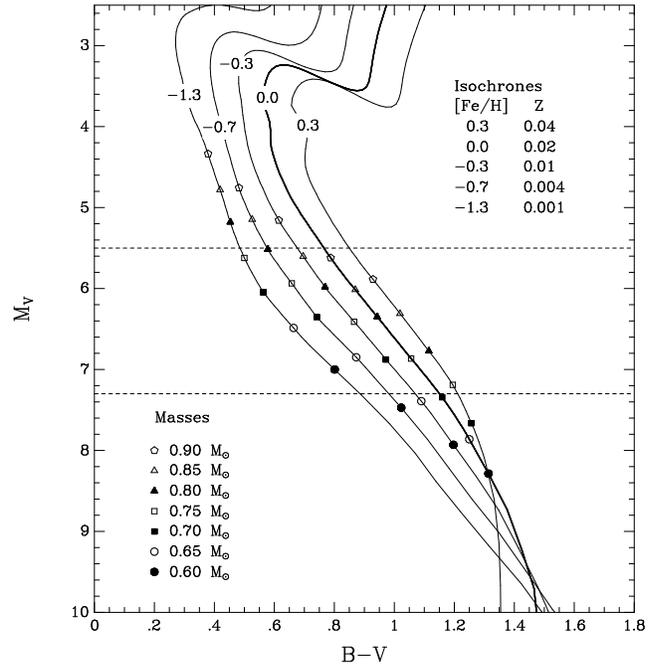,width=84mm}
\caption{Yonsei-Yale (Y$^2$) isochrones for a range of metallicities. The
positions of stars of various mass are marked with different symbols. In the K
dwarf region, $5.5 < M_V < 7.3$ (dashed horizontal lines) there is a simple
relation between mass, absolute magnitude $M_V$ and colour $B-V$.}
\label{YYmass}
\end{figure}

We show in Fig \ref{YYmass} the 5 Gyr Yonsei-Yale (hereafter Y$^2$)
isochrones (Yi et al, 2001) for a range of metallicities.  Also marked are the
positions on each isochrone at which stars of masses 0.90, 0.85, \dots, 0.60
M$_\odot$ lie. The two horizontal lines indicate the absolute magnitude cuts
which were used to construct the basic sample, $5.5 < M_V < 7.3$. Within the
absolute magnitude range of interest, there is clearly a simple relation
between mass and position in the colour-magnitude diagram. We have fit the mass
$M$ (in M$_\odot$) as a function of $M_V$ and $B-V$ as follows

\begin{equation}
M = 0.322\times(B-V) - 0.169\times M_V + 1.551.
\end{equation}

The relation was found by fitting mass, colour and luminosity in the Y$^2$
isochrones for ages ranging between 2 and 10 Gyr. The internal precision in the
relation is quite good, with a scatter of 0.03 M$_\odot$ in the mass
determinations.  We have performed a similar analysis using the Padova
isochrones for three metallicities and an assumed age of 5 Gyr, and find a very
similar fitting relation for mass as a function of $M_V$ and $B-V$ as we did
for the Y$^2$ isochrones.  Masses obtained via the two isochrone sets follow a
1:1 relation closely, but there is an offset of 0.03 \Msun, in the sense that
the Padova masses are lower than the Y$^2$ masses.  This difference is due to
the differences in the adopted helium abundances. The Y$^2$ isochrones assume
$dY/dZ = 2$ whereas the Padova isochrones assume $dY/dZ = 2.25$. The
consequence of this is that the Padova isochrones have more Helium and so
achieve the same luminosities as the Y$^2$ isochrones for slightly lower
masses.  Tests of the isochrones for Hipparcos K dwarfs in binaries and of
known mass (Soederhjelm, 1999) would be of interest to better establish the
scale and zero point of the mass calibration. For our purposes it is sufficient
that the stars are marked by mass on a relative rather than an absolute scale,
as is likely to be the case.

\begin{figure}
\epsfig{file=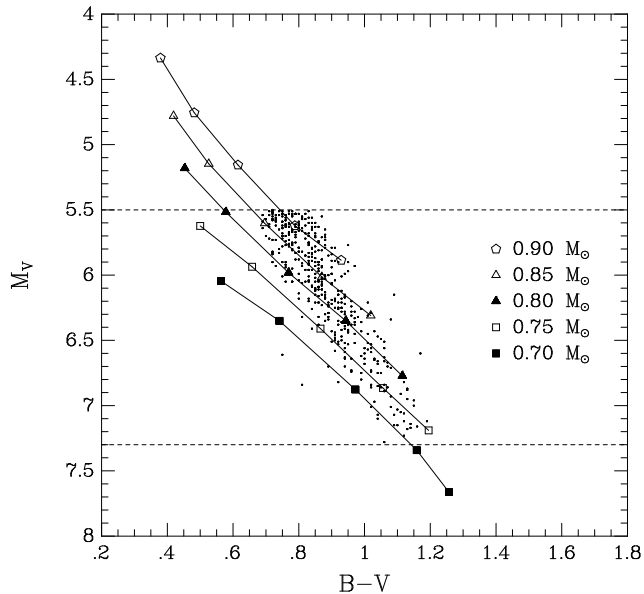,width=84mm}
\caption{Lines of constant mass for the Y$^2$ isochrones, plotted over the K
dwarfs in our basic sample. The dwarfs were initially selected by absolute
magnitude $5.5 < M_V < 7.3$ (shown by dashed horizontal lines). The absolute
magnitude cuts bias the sample by including too many high metallicity stars at
high mass and too many low metallicity stars at low mass. To avoid this we have 
restricted the sample to the mass range $0.75 < $ M/M$_\odot < 0.83$.}
\label{mass}
\end{figure}

Judging from Fig \ref{YYmass}, our dwarfs typically have masses of 0.70-0.90
M$_\odot$. A more detailed plot of the K dwarf region is shown in Fig
\ref{mass}. It is clear that the absolute magnitude cuts bias the sample
(making it broader), by including too many metal rich stars at higher masses
and too many metal poor stars at lower masses (see Fig \ref{MDFcomp}, upper
panel).

\begin{figure}
\epsfig{file=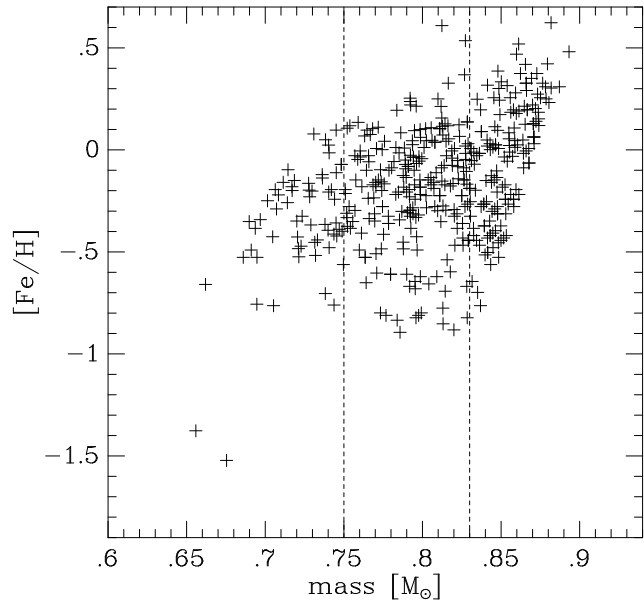,width=84mm}
\caption{Metallicity [Fe/H] versus mass for our K dwarfs. The vertical lines
show the limits within we judge the sample to be substantially complete by
mass, i.e. between 0.75 and 0.83 M$_\odot$. The initial selection by absolute
magnitude (to ensure that stellar evolutionary effects are negligible) causes
the slanted edges to the left and right in the distribution of stars.}
\label{fehmass}
\end{figure}

This is shown in detail in Fig \ref{fehmass}. We have marked on this plot
the mass range in which the sample appears to be complete, i.e. between 0.75
and 0.83 M$_\odot$ (on the scale calibrated to the Y$^2$ isochrones). Selecting
stars within this interval reduces the sample from 431 to 220 K dwarfs.

\begin{figure}
\epsfig{file=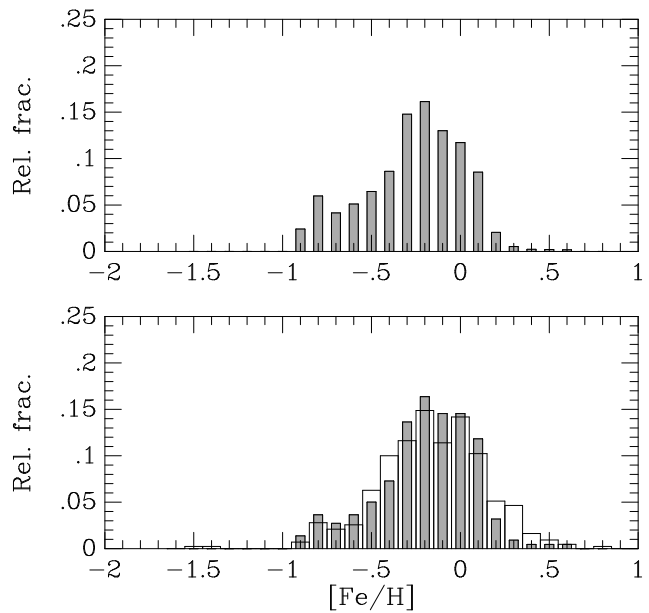,width=84mm}
\caption{Normalised metallicity distribution functions (MDF) for our K
dwarfs. The metallicities are based on [Fe/H]$_{\mathrm KF}$ from Table
\ref{database}. Lower panel : the open histogram is for the basic sample of 431
K dwarfs, selected in the absolute magnitude window $5.5 < M_V < 7.3$. The
filled histogram shows the raw normalised MDF for 220 K dwarfs selected to lie
in the mass range $0.75 < $ M/M$_\odot < 0.83$ (c.f. Fig. \ref{fehmass}). The
main effect of restricting the stars by mass is to remove some stars in the
metal rich tail of the distribution. Upper panel : our final normalised
metallicity distribution of K dwarfs in the vertical column at the Sun
(i.e. including the velocity correction). This histogram is tabulated in Table
\ref{MDFtable}.}
\label{MDF}
\end{figure}

The metallicity distributions for the absolute magnitude limited sample (431
stars) and the mass restricted samples (220 stars) are shown in the lower
panels of Fig \ref{MDF}. The two distributions are quite similar; the main
effect of the mass completeness restriction is to remove stars from the metal
rich tail of the raw MDF.

The upper panel of Fig \ref{MDF} shows our final MDF for the 220 mass
selected K dwarfs after the velocity correction (i.e. by multiplying by the
factor $f$ described in section \ref{shvel} and renormalising).  The velocity
corrections slightly increase the relative fraction of metal weak stars and
slightly decrease the relative fraction of metal rich stars, as expected.  The
MDF is shown in Table \ref{MDFtable}.

\begin{table}
\begin{center}
\caption{Tabulated metallicity distribution function for the final sample of
220 K dwarfs, corrected from local density to coloumn density (via the
metallicity-velocity dispersion relation) and selected in the mass range $0.75
< $ M/M$_\odot < 0.83$. This is the MDF shown in the upper panel of Fig
\ref{MDF}.  The metallicities come from [Fe/H]$_{\mathrm KF}$ in Table
\ref{database}. The error in the relative fractions is based on the Poisson
sampling noise in the raw metallicity distribution. }
\begin{tabular}{rccrcc}
\hline
[Fe/H]  & Rel. Frac & Error& [Fe/H]  & Rel. Frac & Error\\
\hline
$-1.05$ &  0.0000 &  -    & $-0.15$ &  0.1304 & 0.0230\\
$-0.95$ &  0.0238 & 0.0133& $-0.05$ &  0.1178 & 0.0208\\
$-0.85$ &  0.0590 & 0.0206& $ 0.05$ &  0.0860 & 0.0168\\
$-0.75$ &  0.0411 & 0.0165& $ 0.15$ &  0.0207 & 0.0077\\
$-0.65$ &  0.0507 & 0.0177& $ 0.25$ &  0.0052 & 0.0035\\
$-0.55$ &  0.0643 & 0.0192& $ 0.35$ &  0.0023 & 0.0021\\
$-0.45$ &  0.0859 & 0.0213& $ 0.45$ &  0.0020 & 0.0018\\
$-0.35$ &  0.1475 & 0.0268& $ 0.55$ &  0.0017 & 0.0016\\
$-0.25$ &  0.1615 & 0.0268& $ 0.65$ &  0.0000 & -     \\
\hline		                            
\end{tabular}
\label{MDFtable}
\end{center}
\end{table} 

\section{Galactic chemical evolution and K dwarfs}
 
\subsection{The theoretical model}

The model of Galactic chemical evolution we adopt here is that of Chiappini et
al (1997) subsequently modified in Chiappini et al (2001), where a detailed
description can be found. We review here the main ingredients of this model:
\medskip
\begin{itemize}
 
\item The model assumes that the halo + thick disc and the thin disc are
formed during two different infall episodes. The thin disc does not form out of
gas shed from the halo and the thick disc, but simply out of external gas.
Such an interpretation is supported by recent dynamical and kinematical studies
of stars in the outer Galactic halo by Sommer-Larsen et al (1997).  Under these
hypotheses, the infall rate is
 
\begin{eqnarray}
{dG_i(r,t) \over dt} & = & {A(r)\times{(X_{inf})_{i} e^{-t/\tau_T} \over 
\sigma_{Tot}(r,t_{G})}} ~~+ \nonumber \\ 
& & {B(r)\times{(X_{inf})_{i} e^{-(t-t_{max})/\tau_D(r)} \over \sigma_{Tot}(r,t_{G})}}
\label{AB}
\end{eqnarray}
 
\noindent where $\tau_T$ represents the time scale for the formation of the
halo and the thick disc, and $\tau_D(r)$ represents the time scale for disc
formation, which is assumed to increase with Galactocentric distance.  The
best-fit model of Chiappini et al (2001) suggests that the timescale for the
formation of the halo and thick disc is quite short and lie in the range $\sim
0.5-1.0$ Gyr, whereas the timescale for the formation of the thin disc is quite
long ($\sim 7.0$ Gyr for the solar vicinity). This timescale for the thin disc
ensures a very good fit of the new data on the G-dwarf metallicity distribution
(Wyse and Gilmore, 1995; Rocha-Pinto and Maciel, 1996). $A(r)$ and $B(r)$ are
derived by the condition of reproducing the present total surface mass density
distribution in the solar vicinity.  In other words, we integrate Eqn. \ref{AB}
over time up to the present and normalise the left hand side by imposing that
it is consistent with the disc's present total surface mass
density. $(X_{inf})_{i}$ is the abundance of the element $i$ in the infalling
material, $T_G$ the age of the Galaxy, assumed to be 14 Gyr and $t_{max}$ is
the time of maximum gas accretion onto the disc coincident with the end of the
halo-thick disc phase.

\item The Galactic thin disc is approximated by several independent rings, 2
kpc wide, without exchange of matter between them.  Continuous infall of gas
ensures the temporal increase of the surface mass density $\sigma_{Tot}$ in
each ring.
 
\item The instantaneous recycling approximation is relaxed.  This is of
fundamental importance in treating those isotopes, such as $^{14}$N and
$^{56}$Fe, which are mostly produced by long-lived stars.
 
\item The prescription for the star formation rate (SFR) is:

\begin{equation}
SFR \propto \sigma_{Tot}^{k_{2}} \sigma_g^{k_{1}} 
\end{equation}
 
\noindent where $\sigma_{Tot}$ is the total surface mass density and $\sigma_g$
is the surface gas density and $k_{1}=1.5$ and $k_{2}=0.5$.  A threshold in the
surface gas density is also assumed; when the gas density drops below this
threshold the star formation stops.  Two different values for the gas density
threshold are assumed for the halo-thick disc ($\sim 4~ M_{\odot}pc^{-2}$) and
the thin-disc ($\sim 7~ M_{\odot}pc^{-2}$) phases.  The existence of such a
threshold has been suggested by star formation studies (Kennicutt, 1989).

\item For the initial mass function (IMF) we adopt the prescriptions of Scalo
(1986).
 
\item The contributions to the chemical enrichment from supernovae of different
type (Ia, Ib and II) as well as from stars dying as C-O white dwarfs and
contributing processed and unprocessed elements through stellar winds and the
planetary nebula phase are taken into account in great detail (see Chiappini et
al, 2001).
 
\item The adopted nucleosynthesis prescriptions are from: (a) van den Hoek and
Groenewegen (1997) for low and intermediate stellar masses and (b) Thielemann
et al (1993) and Nomoto et al (1997) for SNe Ia. For the massive stars we
considered the yields of Woosley and Weaver (1995) (case B) which include
explosive nucleosynthesis.
\end{itemize}

\subsection{Observational and Cosmic scatter}\label{cosscatter}

The model computes the metallicity distribution function (MDF) assuming no
observational or cosmic scatter. The observational scatter in the metallicities
for the K dwarfs is 0.1 dex (see section \ref{metscatt}). This is quite small
relative to the width of the observed MDF; the model can be easily convolved to
take this into account when comparing to the data. 

Potentially more important is the cosmic, or intrinsic scatter, in the
metallicities of co-eval stars. There is unfortunately no clear consensus
presently regarding the amount of cosmic scatter. 

Edvardsson et al. (1993) found that the age-metallicity relation for F and G
dwarf stars in the solar neighborhood shows a scatter of order 0.2 dex, which
was larger than expected considering the uncertainties in metallicities and
ages. This is of the order the width of the MDF, i.e. most of the MDF is
accounted for by cosmic scatter, and the increase in the mean metallicity as a
function of time plays a minor role.

However, as discussed by Garnett and Kobulnicky (2000) the large scatter found
for the stars in the solar vicinity is inconsistent with the abundance
measurements in nearby spiral and irregular galaxies (e.g. Kobulnicky and
Skillman, 1996) and in the local ISM (Meyer, Jura and Cardelli, 1998), which
show that dispersions in ISM abundances are rather small on kiloparsec scales
or less.  By reanalyzing the Edvardsson et al. sample together with Hipparcos
parallaxes and new age estimates, Garnett and Kobulnicky (2000) found that the
scatter in the age-metallicity relation depends on the distance to the stars in
the sample. They concluded that the intrinsic dispersion in metallicity at
fixed age is less than 0.15 dex for field stars in the solar neighbourhood,
which is closer to the estimate of less than 0.1 dex for Galactic open star
clusters and the ISM (Twarog et al. 1997).

Feltzing, Holmberg and Hurley (2001) argue also this view from a sample 5828
Hipparcos stars for which they derive metallicities and ages; they find a large
scatter in metallicity at any given age and even question the existance of the
Age-Metallicity Relation itself. Rocha-Pinto et al (2000) argue for a very
small cosmic scatter, $\approx 0.1$ dex, based on the age-metallicity relation
they present for nearby stars.

As is well known (Pagel and Tautvaisiene, 1995) the metallicity distribution
function is a very poor test of the Age-Metallicity relation, so we cannot
resolve this issue here. We note that the AMR is less robust than the MDF.  In
the case of the MDF both quantities are directly observable (number of stars
and metallicity), while in the case of the AMR the age scatter contributes
significantly to the apparent metallicity scatter and is not easy to quantify.
We will take the view that the amount of cosmic scatter is uncertain and we
will keep it as a free parameter in the model fitting.

\subsection{Model comparison with the K dwarfs}

\begin{figure}
\epsfig{file=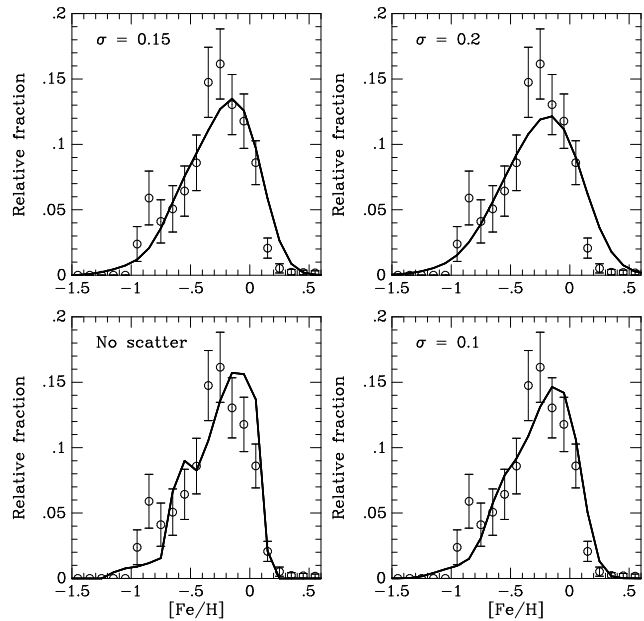,width=84mm}
\caption{Comparison of the model predictions and the metallicity distribution
for K dwarfs. In all panels the data are shown by circles. The lower left panel
shows the two models, unconvolved, as discussed in the text. The remaining
panels show the model convolved with a Gaussian representing the amount of
observatinal scatter and intrinsic scatter.}
\label{modelcomp}
\end{figure}

The metallicity distribution function for the model is compared to the data in
Fig \ref{modelcomp}. The model is a good fit. The data are shown in all
panels by circles and with error bars based on the Poissonian sampling error
(for the 220 star sample).  The solid line shows the model results based on the
Woosley and Weaver (1995) and van den Hoek and Groenewegen (1997) yields. This
model is normalized to the value of the metallicity after $14.0-4.5$ Gyr
(i.e. when the Sun was born).  In the lower left panel of Fig
\ref{modelcomp} the curves show the unconvolved model. In the remaining panels
the model is convolved by a Gaussian representing the observational and cosmic
scatter. The effect of the observational scatter alone is shown in the lower
right panel (0.1 dex). The upper left panel shows the convolved model for a
total scatter of 0.15 dex, equivalent to 0.1 dex observational scatter and 0.11
dex of cosmic scatter, and the upper right panel shows a total scatter of 0.2
dex, equivalent to 0.1 dex observational scatter and 0.17 dex of cosmic
scatter. Clearly, for this model, we cannot constrain the cosmic scatter beyond
noting that it is less than 0.2 dex, and probably less than 0.15 dex. This is
consistent with existing direct observational constraints (section
\ref{cosscatter}. We conclude that the agreement of the model prediction with
the observed K-dwarf distribution is excellent, for any reasonable adopted
cosmic scatter. An independent analysis of the width of the main sequence in
the Hipparcos colour magnitude diagram by Girardi (2002, private communication)
has also obtained the result that the cosmic abundance scatter is probably less
than 0.2 dex.

\section{Summary}

We have calibrated several photometric abundance indices for K dwarfs based on
a sample of 34 G and K dwarfs with accurate, spectroscopically determined
metallicities. Two of the indices use Cousins $R -I$ photometry to estimate
stellar effective temperature and the Geneva $b_1$ or Str{\"o}mgren $m_1$ and
$c_1$ colours.  These indices give metallicity estimates of $\approx 0.2$ dex
accuracy. A third metallicity index uses $B-V$ and $V-I$ broadband colours. A
fourth metallicity index, described in detail in a companion paper (Kotoneva,
Flynn and Jimenez 2002, paper II), is based on absolute magnitude $M_V$ and
$B-V$ colour. These latter two indices appear to provide metallicities of
$\approx 0.1$ dex accuracy.

For a set of newly acquired observations of K dwarfs, we use one of these
indices to obtain the metallicity distribution of a near complete sample of K
dwarfs in the Solar neighbourhood drawn from the Hipparcos catalog. Care has
been taken to remove the multiple stars from the sample, for which
metallicities cannot be measured accurately. Through isochrones we assign
masses to the K dwarfs, and select those K dwarfs which fall into a mass window
$0.75 < $ M/M$_\odot < 0.83$, within which the sample is near to complete. This
yields a sample of 220 K dwarfs.

The metallicity distribution of the 220 K dwarfs is strongly peaked near the
solar metallicity, confirming the existence of the ``G dwarf problem'' amongst
K dwarfs, as seen in several earlier studies (Marsakov and Shevelev (1988),
Flynn and Morell (1997) and Rocha-Pinto and Maciel (1998).  We compare the
metallicity distribution with Galactic chemical evolution models of Chiappini
et al (2001). In the context of this model of the metallicity evolution, we
find that the amount of cosmic scatter in the metallicities is small, not more
than 0.15 dex. The model match the data well, indicating that the disc was
formed via infall processes over an extended time scale of order 7 Gyr.


\section*{Acknowledgments}

This research was supported by the Academy of Finland, the Jenny and Antti
Wihuri Foundation, the Magnus Ehrnrooth Foundation, the Emil Aaltonen
Foundation. EK thanks Prof.~Robert Shobbrook for his valuable comments and help
at Siding Spring Observatory. We thank Leena T{\"a}htinen for help with the
observations taken at La Palma, Johan Holmberg for assistance with the
Hipparcos and Tycho catalogs and cheerfully rendering much programming advice,
Helio Rocha-Pinto for his help in computing the chrosmopheric corrections and
all his helpful comments. Bernard Pagel, Raul Jimenez, Brad Gibson, Yeshe
Fenner and Leo Girardi made many insightful comments. EK thanks the
Astronomical Observatory of Trieste for its hospitality during a one month
visit, where part of this work was carried out.

\end{document}